# A Novel Low Power Non-Volatile SRAM Cell with Self Write Termination

Kanika Monga, Akul Malhotra, Nitin Chaturvedi, S. Gurunayaranan
Department of EEE, Birla Institute of Technology and Science, Pilani
Pilani, Rajasthan – 333031, India
p20170027@pilani.bits-pilani.ac.in, f2016171@pilani.bits-pilani.ac.in, nitin80@pilani.bits-pilani.ac.in, sguru@pilani.bits-pilani.ac.in

*Abstract* — A non-volatile SRAM cell is proposed for low power applications using Spin Transfer Torque-Magnetic Tunnel Junction (STT-MTJ) devices. This novel cell offers non-volatile storage, thus allowing selected blocks of SRAM to be switched off during standby operation. To further increase the power savings, a write termination circuit is designed which detects completion of MTJ write and closes the bidirectional current path for the MTJ. A reduction of 25.81% in the number of transistors and reduction of 2.95% in the power consumption is achieved in comparison to prior work on write termination circuits.

## I. INTRODUCTION

Aggressive CMOS scaling have resulted in increased leakage currents in the deep-sub micrometer regions which has become one of the critical problem for future microprocessors. Leakage currents lead to enormous standby power and memory degradation. Several leakage current reduction techniques have been proposed [1] to mitigate this problem. However, introduction of non-volatility to the memory cell can be a promising solution to overcome the problem of high standby power. The non-volatile memory elements added to the SRAM cell will store the logic state information. Since the state information is stored in the non-volatile medium, the memory block can be completely powered down during standby to eliminate all leakage currents. When the SRAM block is activated, the logic state information will be restored in it from the non-volatile memory elements. Thus, the power consumption of the cell can be reduced without any effect on the operating frequency of the cell.

Of the many nonvolatile memory technologies proposed, Spin Transfer Torque-Magnetic Random Access Memories (STT-MRAM) are most suitable for on-chip cache applications [2]. STT-MRAM can achieve the integration density of DRAM [3] and potentially match the performance of SRAM. Compared to other competing nonvolatile memory technologies, STT-MRAM is compatible with the CMOS fabrication process and has a better combination of retention time, endurance, and reliability [4]. As a storage element, an STT-MRAM cell consists of a magnetic tunnel junction (MTJ) which is mainly composed of one ultra-thin oxide tunneling barrier (TB) layer (e.g., MgO) sandwiched by two ferromagnetic (FM) layers (e.g., CoFeB) [5]. In general, the magnetization orientation of one FM layer is fixed in one direction, called pinned layer, whereas the magnetization of the other FM layer, termed free layer, is reversible. The relative magnetization orientation of the two layers determines resistance state of the MTJ: Low resistance (parallel state) and high resistance (anti-parallel state). For writing into the STT-MTJ, bidirectional current larger than a threshold current value, called the critical current, $I_{C0}$, is required [6], [7].

Although STT switching is advantageous in terms of its simple switching method, it has many reliability issues due to its intrinsic stochasticity [8]. This means that the duration of a switching event is not constant and varies from one event to another. Thus, a MTJ write operation of fixed duration can result in write failures. To ensure that no write errors occur, the duration of the MTJ write operation is increased. However, this results in a higher power consumption [9] as the MTJ may switch earlier than the write pulse duration. One possible solution to reduce write energy is to use additional circuitry which senses the completion of MTJ write operation and terminates it immediately.

In this paper, a novel STT-MRAM based non-volatile SRAM (nv-sram) cell with write termination is proposed. The proposed cell has 4 operations associated with it: A bit can be written into the cell (write operation), a bit can be read from the cell (read operation), the bit stored in the SRAM cell can be written into the non-volatile storage element (MTJ) (backup operation/MTJ write) and the bit stored in the MTJ's can be stored back into the SRAM cell when it resumes operation (restore operation). Thus, the proposed nv-SRAM cell functions as a regular SRAM cell during its active period. However, the cell is capable of storing bit data into the MTJ's during power off state, resulting in tremendous standby power savings. The proposed cell also contains a write completion detection and termination circuit, which improves the power consumption during the backup operation. It should be noted that normal SRAM write operation and MTJ write operation are completely different operation and proposed write termination circuit is used for MTJ write only.

The rest of the paper is organized as follows. Section II provides the motivation behind the work. In Section III, existing techniques for write power reduction is presented. Section IV proposes the non-volatile SRAM cell with write termination circuit. Section V discusses the simulation results and power savings due to the proposed write termination circuitry. It also presents comparison result of proposed write termination circuit with other write termination circuits. Finally, Section VI concludes the paper.

## II. MOTIVATION

One of the possible approaches to reduce the standby power is to turn off the memory block when not in use. Consequently, state of the system must be stored and recalled before power off and after power on respectively. The conventional way of checkpointing involves storage of state to off-chip non-volatile memory which results in significant energy and latency overhead. Therefore, this paper proposes MTJ based hybrid non-volatile SRAM cell which can retains the state on-chip thereby reduces the power consumption and store/restore speed. However, one critical issue with the MTJ is its high write energy. Addressing this issue, the proposed nv-sram cell is augmented with a termination circuit that detect the successful switching of MTJ states and terminates the write current to save power.

## III. RELATED WORK

To address the challenges of high write energy various attempts have been made in the past. At the device level, S. Wolf et. al [10] proposed perpendicular magnetic anisotropy MTJ which had the significant lower write energy than in-plane anisotropy MTJ. C. Smullen et. al [11] proposed method to improve switching efficiency by reducing retention time. However this method is suitable only for application with short idle periods. Application with long idle period requires peroidic refresh which again result in higher power consumption. At circuit level, G. Sun et. al [12] proposed the method of read before write to terminate the write operation in case of redundant write. However, read operation itself consumes energy and it also does not consider asymmetric and stochastic nature of write operation. To account for asymmetricity and stochasticity of MTJ write D. Suzuki et. al [13] proposed 1T-1MTJ structure with write termination circuit. The voltage variation on node SL and BL is monitored to detect the switching and terminate the write operation. M. Gupta et. al [14] modified the circuit proposed by D. Suzuki to reduce the number of transistors. In this paper a hybrid non-volatile SRAM cell is proposed which uses similar concept of continuous node voltage monitoring to terminate the write operation with less number of transistor. As soon as writing into the MTJ is complete, voltage on node (connected to pinned layer of MTJ) rises due to change in resistance of MTJ which serves as a detection signal for successful write operation. This detection signal is then used to disable the write path, thereby terminating the write operation.

## IV. PROPOSED NV-SRAM CELL

Fig. 1 shows the schematic of the proposed STT-MTJ based non-volatile SRAM cell. In this work, the stored data '1' corresponds to the low resistance state (parallel state) of MTJ1 and the high resistance state (anti-parallel state) of MTJ2. Similarly, stored data '0' corresponds to the high resistance state (anti-parallel state) of MTJ1 and the low resistance state (parallel state) of MTJ2. The circuit has four operations: write, read, backup and restore.

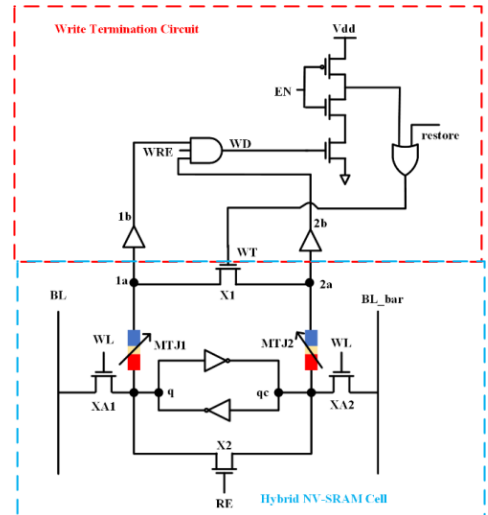

Fig 1. Proposed non-volatile SRAM cell with MTJ write termination

### A. Read and Write operation

The cell functions as a standard SRAM cell during this mode of operation as shown in Fig 2. For writing data into the SRAM cell, the word line (WL) is asserted high. Depending on the value of BL and BL_bar, one of the nodes of the cross-coupled inverters gets charged and the other discharged. Thus, the bit value is written into the SRAM cell. To read from the SRAM cell, the bitlines BL and BL_bar are precharged to Vdd. The bitline corresponding to the node of the cross coupled inverters which has the value 0 starts to discharge. The drop in voltage is sensed by a sense amplifier. Thus, the bit value stored in the SRAM cell is read.

### B. Backup operation

In this mode of operation, value present at nodes q and qc is written into MTJ1 and MTJ2 respectively. Therefore, this operation is also called as MTJ write operation. The backup operation is initiated with the control signal WRE asserted high and the control signal EN asserted low (for a very short duration) as shown in Fig 2. The control signal WRE remains at Vdd for entire backup operation, which is a duration long enough for successful switching of MTJ states. The low EN pulse charges WT to Vdd, which opens the equalization transistor X1 between the two MTJ's. Thus, a current path is established for both the MTJ's. However, the direction of the current is determined by the value of q and qc.

While writing into the MTJ two cases can arise: 1) the bit value to be written and the current states of the MTJs are the same, or 2) the bit value to be written and the current states of the MTJs are different. Let us first consider the case when both are same. The nodes 1a and 2a attain a value greater than the threshold value of buffers therefore both the nodes 1b and 2b rises to Vdd. Now since WRE is already high, WD attain the value Vdd. As soon as EN rises back to Vdd, WT falls to 0, closing the equalization transistor X1 and immediately terminating the current flow in the MTJ's.

However, when the bit value to be written and the current states of the MTJ's are different as shown in Fig 2, the nodes 1a and 2a initially attain a value lower than the threshold value of the buffers which results in nodes 1b and 2b falling to 0 and hence WD also 0. On the other hand, the continuous current flow results in change of MTJ states, thus changing their resistance values. Due to the change in resistances, the voltage values at nodes 1a and 2a start rising. The change in voltage is used to signify the completion of writing into the MTJ. Once the node voltages 1a and 2a cross the threshold voltage for the buffers, WD rises to Vdd, followed by discharging of WT and closing the equalization transistor X1. As soon as the path between the two MTJs is closed, node 1a takes the value of node q and node 2a takes the value of node qc. This makes WD go back to 0V. Since EN is at Vdd, WT remains at 0V and the current path for both the MTJ is broken. This results in termination of MTJ write current and contributes to power saving.

*C. Restore operation*

The restore operation is initiated by asserting control signal RE high as shown in Fig 2. This turns on the equalization transistor X2 and bring nodes q and qc to the same voltage value. Once control signal RE falls back to 0, the restore signal rises to Vdd. This is done to turn on the equalization transistor X1, opening up a path between the two MTJ. As soon as, the control signal RE falls back to 0 and restore rises to Vdd, a voltage difference appears between q and qc. Due to this voltage difference, current flows from the node that has the low resistance state MTJ connected to it to the node that has the high resistance state MTJ connected to it. Thus, if $q$ was connected to the MTJ that was in low resistance (parallel) state, q would rise to a higher voltage value than qc. The differential voltage between the nodes would be amplified to full swing by the SRAM cell, restoring the bit values at nodes q and qc. Once restore signal falls back to 0, transistor X1 would be turned OFF, thus stopping the current flow between q and qc and finishing the restore operation.

## V. SIMULATION RESULTS AND COMPARISON

Simulations have been performed using a physics-based STT-MTJ compact model [15] and a FinFET 20 nm design kit, to demonstrate the functionality of the proposed nv-sram cell with write termination. Table I shows the major parameters of STT-MTJ in the simulations.

TABLE I: PARAMETERS OF MTJ

| Parameters | Values |
|---|---|
| Free layer Volume (W(nm)*L(nm)*T(nm)) | 20*20*1.4 |
| Oxide layer Thickness (nm) | 1.15 |
| Saturation Magnetization (emu/cm$^3$) | 700 |
| Damping Factor | 0.028 |
| Rp (Low Resistance) | 5.3-5.7 KΩ |
| Rap (High Resistance) (KΩ) | 10.2-15.03 |
| TMR ratio | >100% |
| Io (Switching Critical Current) | 27uA |
| Ea (Thermal Activation Energy) | 56*KB*T |
| Temperature | 300K |

Table II shows the performance analysis of the proposed nv-sram cell with write termination compared to the proposed nv-sram cell without write termination. It also compares the proposed nv-sram cell with write termination to state of art write termination circuit proposed in [13] and [14]. The parameters used for this comparison are number of transistors in the write termination circuitry, power consumption of writing bit '0', power consumption for writing bit '1', mean write energy and power savings measured with respect to the proposed nv-sram cell without write termination.

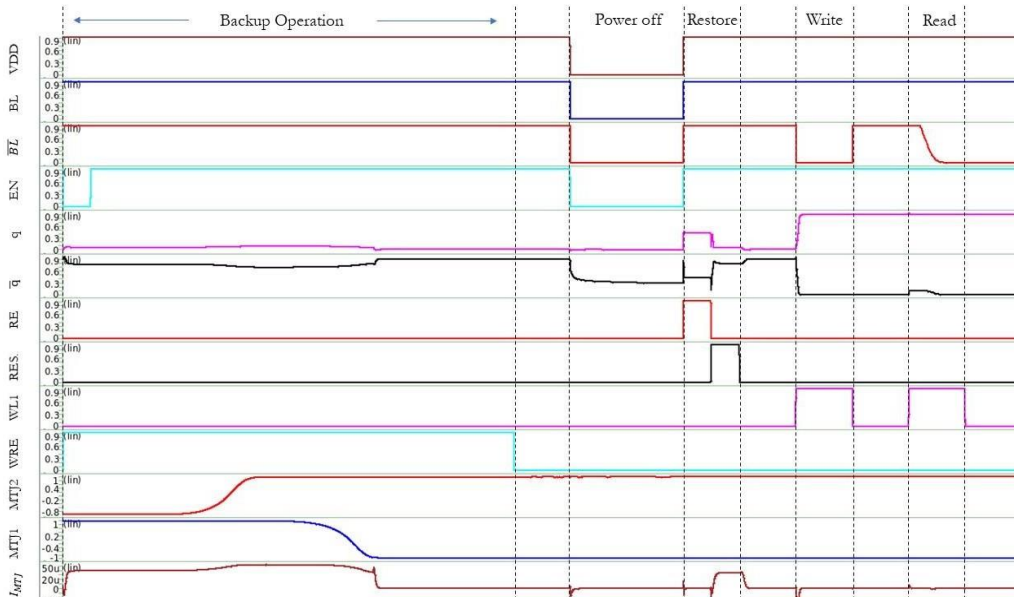

Fig 2. Simulation result showing the four operations and the various signals associated with them

TABLE II: PERFORMANCE COMPARISON OF WRITE TERMINATION CIRCUITS

|  | Proposed nv-sram cell without write termination | [13] | [14] | Proposed nv-sram cell with write termination |
|---|---|---|---|---|
| #of transistor* | - | 40 | 31 | 23 |
| Write power for '0' [pJ/bit] | 0.313 | 0.247 | 0.230 | 0.257 |
| Write power for '1' [pJ/bit] | 0.313 | 0.150 | 0.150 | 0.257 |
| Mean write energy [pJ/bit] | 0.313 | 0.198 | 0.190 | 0.257 |
| Power savings during backup operation (%) | - | 11.75 | 14.95 | 17.88 |

* Only the transistors in the write termination circuit are counted.

From the Table II it is observed that reduction of 42.50% and 25.81% in number of transistors in achieved with the proposed write termination circuit compared to termination circuit of [13] and [14] respectively. The proposed write termination circuit is able to save 17.88 % power when compared with cell without write termination circuit. Also, power savings of 6.15% and 2.95% is achieved by the write termination circuitry in the proposed nv-sram cell with write termination in comparison to [13] and [14].

## VI. CONCLUSION

In this paper, we presented a low power non-volatile SRAM cell with self-write termination. Simulations using STT-MTJ and 20 nm FinFET models have verified its performance and functionality. It is shown that the proposed cell could perform functions: read, write, backup and restore, successfully. Despite several advantages STT-MTJ suffers from large switching latency and high switching energy. Therefore, a novel write termination circuit is proposed in this work to reduce the MTJ write energy during the backup operation. The MTJ write termination circuit achieved a power reduction of 17.88%, which is 6.95% and 2.95% more than what prior works [12] and [13] respectively.